\documentclass[aip,jap,twocolumn, reprint]{revtex4-1}

\usepackage{amsmath}
\usepackage{graphicx}
\usepackage{units}
\usepackage{color}

\begin{document}

\title{Accessing defect dynamics using intense, nanosecond pulsed ion beams}
\author{A. Persaud}
\email{APersaud@lbl.gov}
\affiliation{Accelerator and Fusion Research Division, Lawrence Berkeley National Laboratory, Berkeley, CA 94720, USA}

\author{J. J. Barnard}
\affiliation{Lawrence Livermore National Laboratory, Livermore, CA 94550, USA}
\author{H. Guo}
\affiliation{Materials Sciences Division, Lawrence Berkeley National Laboratory, Berkeley, CA 94720, USA}
\author{P. Hosemann}
\affiliation{Materials Sciences Division, Lawrence Berkeley National Laboratory, Berkeley, CA 94720, USA}
\affiliation{Nuclear Engineering Department, University of California, Berkeley, CA 94720, USA}
\author{S. Lidia}
\affiliation{Accelerator and Fusion Research Division, Lawrence Berkeley National Laboratory, Berkeley, CA 94720, USA}
\author{A. M. Minor}
\affiliation{Materials Sciences Division, Lawrence Berkeley National Laboratory, Berkeley, CA 94720, USA}
\affiliation{Department of Materials Science and Engineering, University of California, Berkeley, CA 94720, USA}
\author{P. A. Seidl}
\affiliation{Accelerator and Fusion Research Division, Lawrence Berkeley National Laboratory, Berkeley, CA 94720, USA}
\author{T. Schenkel}
\affiliation{Accelerator and Fusion Research Division, Lawrence Berkeley National Laboratory, Berkeley, CA 94720, USA}

\begin{abstract}
  Gaining \textit{in-situ} access to relaxation dynamics of radiation induced
  defects will lead to a better understanding of materials and is
  important for the verification of theoretical models and
  simulations. We show preliminary results from experiments at the new
  Neutralized Drift Compression Experiment (NDCX-II) at Lawrence
  Berkeley National Laboratory that will enable \textit{in-situ} access to
  defect dynamics through pump-probe experiments. Here, the unique
  capabilities of the NDCX-II accelerator to generate intense,
  nanosecond pulsed ion beams are utilized. Preliminary data of
  channeling experiments using lithium and potassium ions and silicon
  membranes are shown. We compare these data to simulation results
  using Crystal Trim. Furthermore, we discuss the improvements to the
  accelerator to higher performance levels and the new diagnostics
  tools that are being incorporated.
\end{abstract}

\maketitle

\section{Introduction}

Defects in materials created by ion irradiation present a multi-scale
problem with defect lifetimes ranging from picoseconds to years, see
\citet{DiazdelaRubia2000} and \citet{Victoria2000}. Most of the defects
self-anneal with a time scale of the order of picoseconds, see
\citet{Bai2010}. Currently, simulation tools can only assume certain
time constants since experimental verification for fast process on the
time scale of picoseconds has not been possible, see
\citet{Stuchbery1999} and \citet{Myers2012}. However, access to these
processes is important for a better understanding of materials that
play an important role in the development of radiation hard
electronics and the development for advanced materials for future
fusion reactors and the next generation of nuclear reactors, see
\citet{Zinkle2013} and \citet{ZinkleWas2013}. At the new Neutralized
Drift Compression Experiment (NDCX-II) short, intense beam pulses are
available that can provide a pump pulse for pump-probe experiment, see
\citet{Friedman2009}, \citet{Waldron2014}, and \citet{Schenkel2013}. We
report on experimental results using beams of lithium and potassium
ions to probe crystalline silicon samples. We present data from
channeling experiments where the ion beam pulse is also used as a
diagnostic tool. The mechanism exploited in these measurements is
based on the channeling effect. Ions that channel have a longer range
and lower energy loss in the material. The possibility of a channel
event is strongly dependent on the crystalline integrity, or the
amount of damage present in the material. A non-channeling ion hitting
the material will create many interstitial-vacancy pairs in the
material, temporarily blocking many channels in the material before
the defects self-anneal. If the fluence of the ions is high enough,
other ions will hit the same area as the remnants of the earlier ion
cascade (before the self-annealing took place) and the channel
transmission will be impeded. Therefore, we expect that the channeling
current will change depending on the fluence, e.g., for low fluences
channeling will be unobstructed (no overlapping cascades) and for high
fluences the channeling current will decrease during a single beam
pulse (many overlapping cascades). The transmitted beam current can be
monitored with high temporal resolution and by restricting the
measured ions to a small scattering angle, mostly channeled ions can
be recorded. Noise from non-channeled ions can be further reduced by
time-of-flight measurement, since channeled ions will have a smaller
energy loss than scattered ions.

\section{Experiment}
The source technology employed at NDCX-II currently allows the use of
lithium and potassium ions.  (Other alkalis, such as sodium and cesium
are also feasible.) The ions are extracted from a \unit[10]{cm} diameter area
over a time of \unit[600]{ns}. The extraction optics focus the beam to a
radius of about \unit[2-3]{cm}. Using an induction type accelerator the ions
are then transported along a \unit[9]{m} beamline where they are compressed
longitudinally to a full-width-half-maximum (FWHM) beam pulse of about
\unit[20-30]{ns}. The last few solenoids in the accelerator are adjusted to
focus the beam down to a spot size of \unit[5]{mm} radius at the target. The
ion energy of the beam is \unit[135]{keV} for an uncompressed (\unit[600]{ns} long)
beam and \unit[300]{keV} for a compressed beam. The repetition rate of the
accelerator is 2-3 shots per minute.

The peak current in the experiments is presently around \unit[1]{A},
corresponding to $10^{11}$ ions ($\unit[\sim50]{nC}$ for Li and
$\unit[\sim12]{nC}$ for K) that are
delivered per pulse. Ion beam transport with such high peak currents
is dominated by space charge effects, and NDCX-II has been specially
developed to be able to transport beams under these extreme space
charge conditions, see \citet{Friedman2009} and \citet{Waldron2014}.

At the target chamber the beam can be monitored using a scintillator
and a fast camera, as well as a specially designed fast Faraday cup to
monitor the beam current over time. We mount our silicon membranes
(\unit[250]{nm} and \unit[1000]{nm}) at the same location as the scintillator so that
we can optically confirm the beam size. The transmitted beam is then
recorded using a fast Faraday cup positioned \unit[35]{cm} behind the
membrane. This distance increases the sensitivity to channeled ions
(smaller acceptance angle), since scattered ions are preferentially
scattered out of the acceptance angle of the Faraday cup. We can
furthermore utilize the drift distance between membrane and Faraday
cup for time-of-flight measurements.

The target is mounted on a goniometer with x, y and z motion allowing
for a precise alignment of the target to the beam spot recorded using
the camera and scintillator. The goniometer also has the ability to
rotate the sample allowing measurements of the channeling effect for
different incident angles. The sample is mounted on a larger frame so
that only beam passing through the sample can reach the Faraday cup
even at large angles.

Improvements to the accelerator that are presently underway will allow
for a beam energy of \unit[1.2]{MeV}, and \unit[1-2]{ns} long beam pulses with a spot
size of $\unit[\sim1]{mm}$ radius. This will provide up to 400 times increase in
ion fluence. It will be possible following the installation of a
plasma filled drift line at the end of the accelerator.  This
drift-compression section will allow the beam bunch, which has an
ordered head-to-tail velocity ramp, to bunch to $\unit[\sim1]{ns}$ unimpeded by its
own space charge repulsion.  Installation of an \unit[8]{T} solenoid near the
target focuses the beam to $\unit[\sim1]{mm}$, also in the presence of a
neutralizing background plasma.

\section{Results}
\begin{figure}[ht!]
  \centering
  \includegraphics[width=\linewidth]{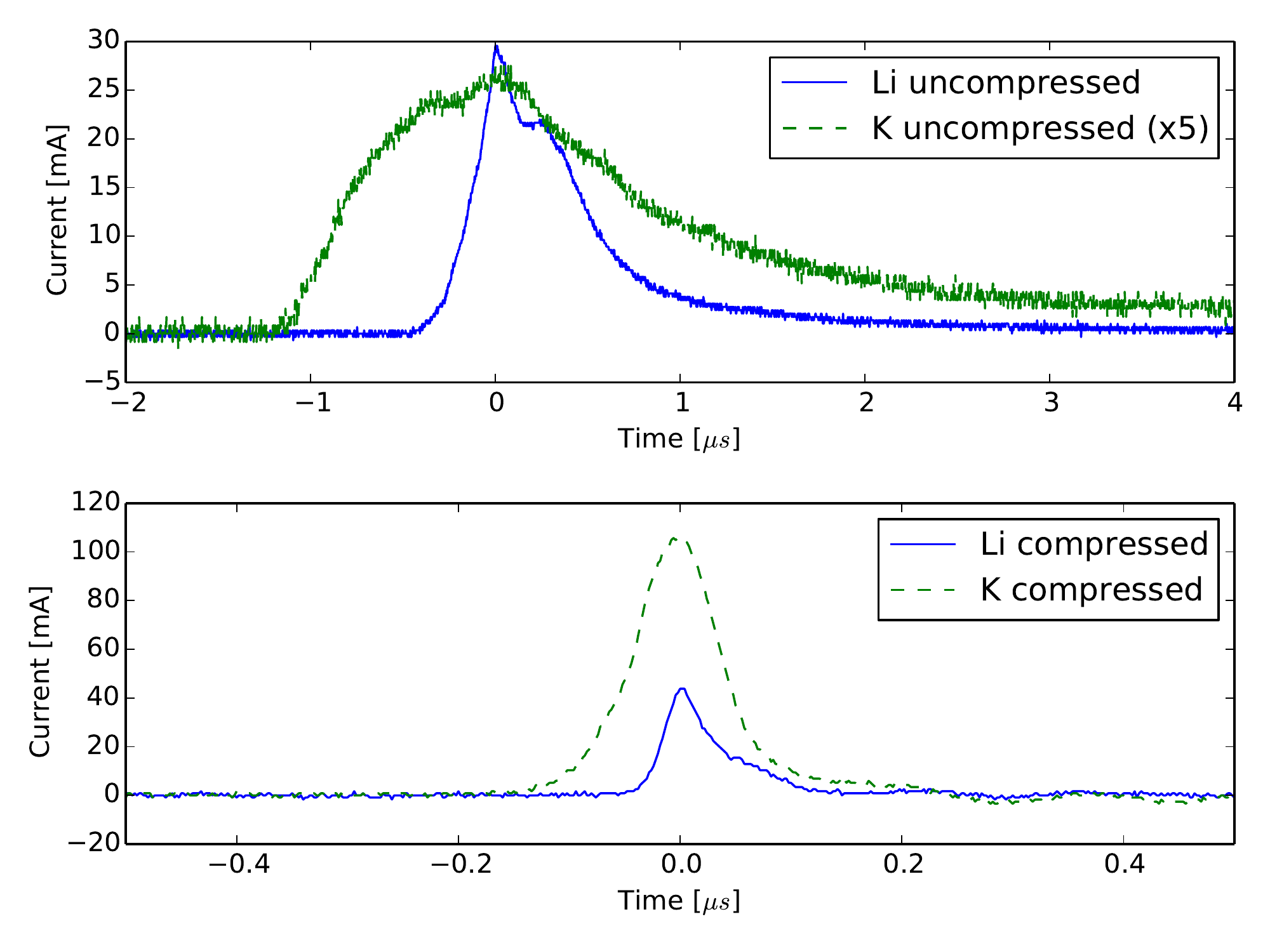}
  \caption{Beam profile measured with a Faraday-cup at the target
    position for compressed and uncompressed beam of lithium and
    potassium.}
  \label{fig:fcup}
\end{figure}

Both the lithium and the potassium sources produce a very reliable
beam with reproducible temporal and spatial profiles for many shots
over periods of days and weeks with minor retuning.
\begin{figure}[ht!]
  \centering
  \includegraphics[width=\linewidth]{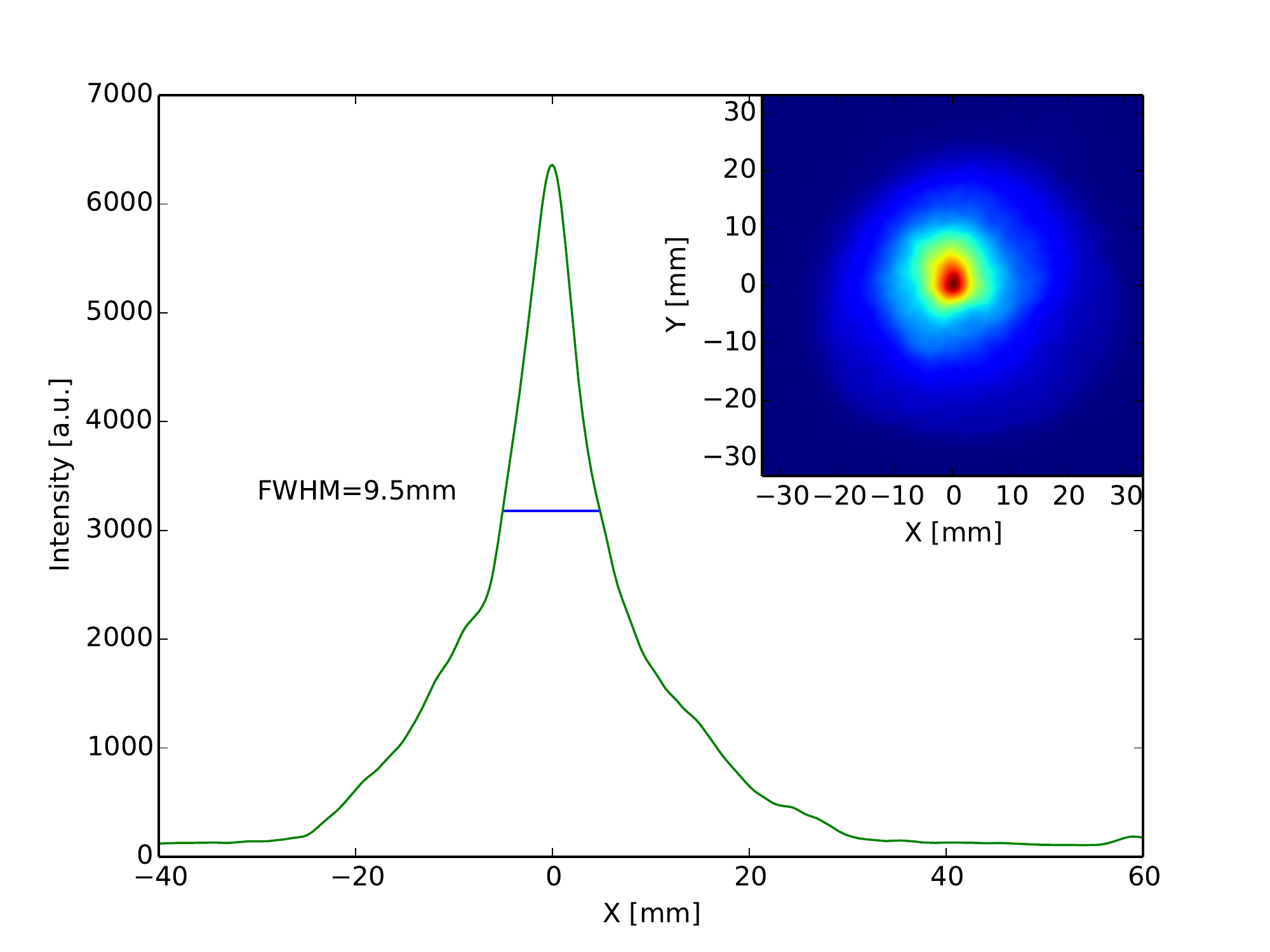}
  \caption{Beam profile of a compressed potassium beam taken with a
    scintillator. The inset shows the scintillator image.}
  \label{fig:k-compressed}
\end{figure}
The beam has a Gaussian transverse profile. Figure~\ref{fig:k-compressed} shows a
scintillator image of a compressed potassium beam.  We find a
shot-to-shot variation of the peak position of \unit[0.2]{mm} (rms) in the
horizontal and vertical direction. Intensity variations are of the
order of 5\% percent and the main source for this is the variation in
temperature at the filament, which currently is not regulated (the
filament is voltage controlled). A feedback loop will be implement in
the future which will create a more stable beam in regard to the
intensity. In Figure~\ref{fig:fcup} a beam profile measured on a Faraday cup at the
target position is shown for a compressed beam and an uncompressed
beam for both lithium and potassium ions.

\begin{figure}[ht!]
  \centering
  \includegraphics[width=\linewidth]{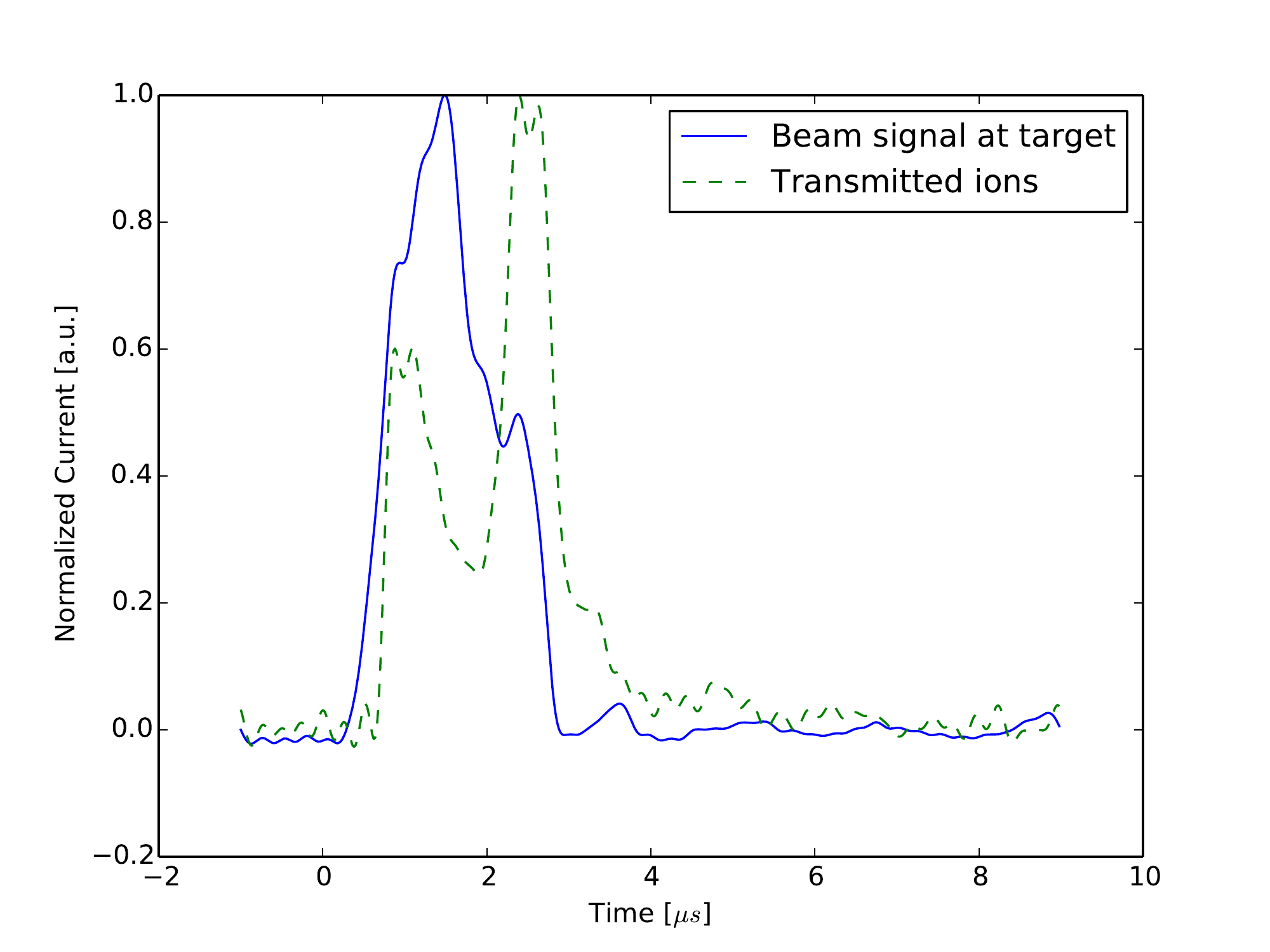}
  \caption{\unit[135]{keV} potassium beam. The blue (solid) curve shows a beam
    pickup measured on the membrane. The green (dashed) curve shows
    the transmitted ions measured in a Faraday-cup \unit[35]{cm} behind the
    membrane.}
  \label{fig:k-membrane}
\end{figure}

\begin{figure}[ht!]
  \centering
  \includegraphics[width=\linewidth]{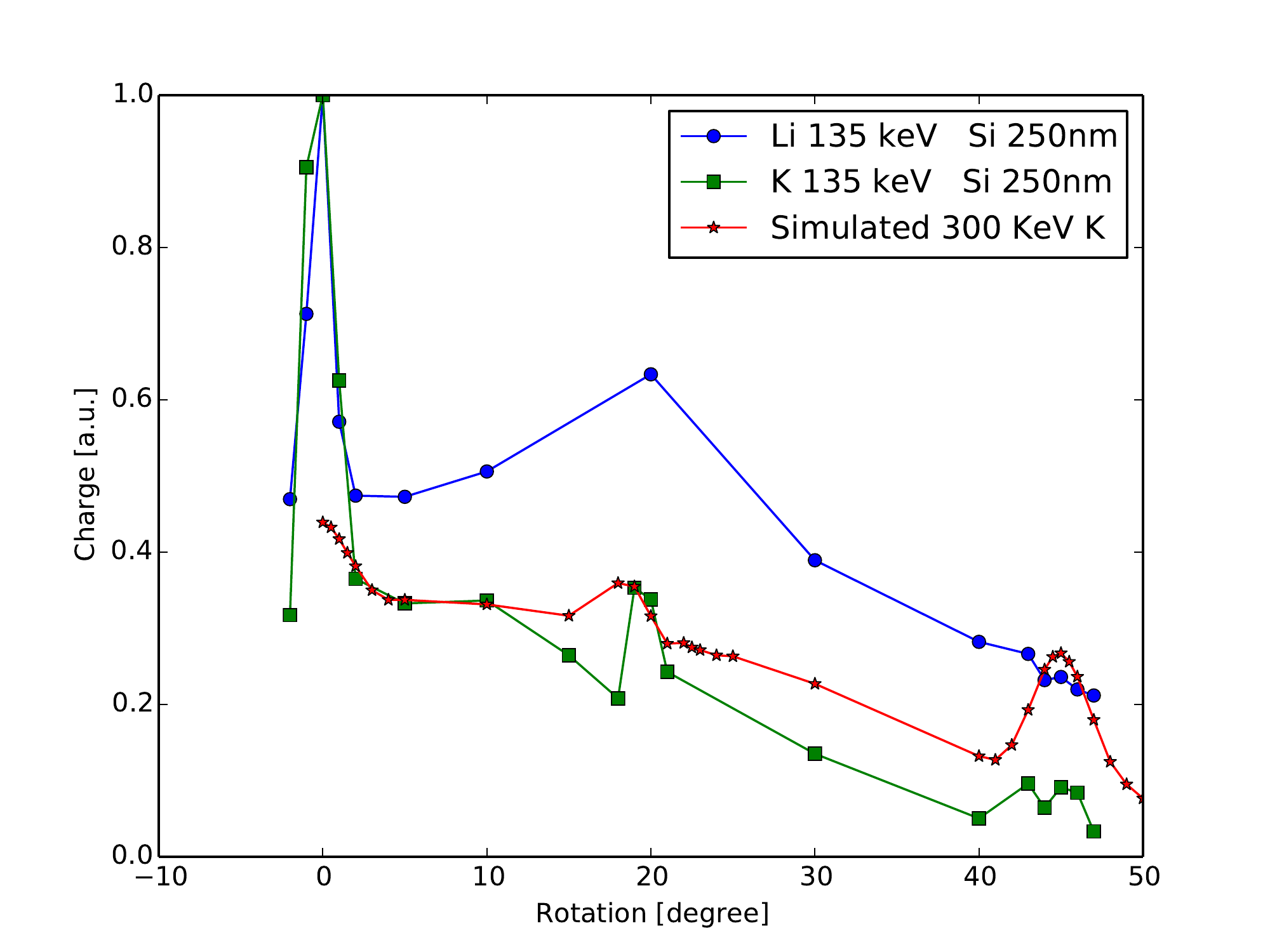}
  \caption{Channeling peaks vs. rotation angle. The effects of the
    well known crystal channel structure can be seen clearly at the
    corresponding angles. The charge for both runs has been
    normalized}
  \label{channeling}
\end{figure}

A beam current waveform for a transmitted potassium beam through a
\unit[250]{nm} thick silicon membrane is shown in
Figure~\ref{fig:k-membrane}. We can integrate the measured current
and plot the transmitted charge versus the rotation angle of the
target membrane. The result is shown in Figure~\ref{channeling}. Channeling peaks can
be clearly seen around 0, 18 and 45 degree, corresponding to the
$\langle$100$\rangle$, $\langle$310$\rangle$ and $\langle$110$\rangle$
channeling directions in the crystal.

The general decrease in amplitude for the non-channeled current can be
attributed to an increased thickness and a smaller effective area due
to the rotation of the membrane.

Figure~\ref{channeling} also shows the results from a Crystal Trim simulation, see
\citet{Posselt1994}, for \unit[300]{keV} K+ ions stopping in crystalline Si (with
\unit[3]{nm} layer of amorphous SiO$_2$ to simulate a native oxide). The
simulation uses a $\unit[1.3\times10^{11}]{cm^{-2}}$ fluence and a beam divergence of 1
degree. The fraction of transmitted ions in Crystal Trim was
calculated by taking the fraction of ions with a range larger than the
membrane thickness of \unit[250]{nm} compared to the total number of ions that
get implanted at all depths. This will underestimate contrast from the
channeling effect in the transmitted ion signal, since our experiment
is set up to only accept ions with a small scattering angle.  For
Figure~\ref{channeling} the simulated data was scaled to fit the 10 degree measured
data point for potassium. Furthermore a cosine dependence of the
exposed area under rotation was taken into account. The simulation
confirms the measured channeling peaks. Further simulations also show
that we can expect changes in the time resolved transmission for
potassium fluences exceeding $\unit[\sim10^{12}]{cm^{-2}}$.

As demonstrated in \citet{Guo2014} for lithium ions, the fluence is
currently not high enough to observe effects from damage build-up due
to overlapping cascades. Potassium ions create more damage due to
their increased mass, but no effect could be observed so far which is
in agreement with the simulations results for the needed fluences. We
will be able to deliver the required higher fluences to samples once
the additional focusing magnet is installed. With faster diagnostics,
e.g. employing streak techniques, we then aim at tracking the
evolution of the channeled ion fraction on a $\unit[\sim10]{ps}$ time scale.

\section{Conclusion}
We have shown that channeling experiments using thin membranes can be
realized with intense, pulsed ion beams from NDCX-II by monitoring
transmitted ion currents and that the concept of detecting defects
dynamics by monitoring channeled ions is viable.  However, with the
current beam current, spot size, and pulse width no effect from damage
build-up could be observed. We attribute this to the fact that we
currently do not have high enough fluences to achieve overlapping
damage cascades during the ion bombardment.

The ongoing upgrade of the accelerator aims at achieving smaller beam
spot sizes (x5 in radius), shorter pulses (x20-30) and higher ion
kinetic energies, up to \unit[1.2]{MeV}. Thus we can expect an increase in
fluence of a factor of x400. We expect to see effects of damage
build-up at short timescales at these fluences as indicated by Crystal
Trim simulations and further drive thin foils to temperatures
approaching \unit[1]{eV}.

Furthermore, we started to integrate new detection capabilities such
as optical detection of defect recombination (ionoluminescence) using
fast photodetectors and spectrometers coupled with a streak
camera. This will allow us to resolve optical changes with a time
resolution of picoseconds. Ion scattering using the channeling effect
will also be implemented in backscattering geometry.  This will allow
us to probe a broader range of targets and remove the limitation to
thin, single crystal membranes, which are currently needed for the
channeling experiments and enable us to probe a much broader range of
materials, such as ceramics for nuclear energy applications or
materials for future fusion reactors, such as tungsten alloys.

\section*{Acknowledgements}
This work was supported by the Office of Science of the U.S. DOE and
by the LDRD Program at Lawrence Berkeley National Laboratory under
contract no. DE-AC02-05CH11231. AM was supported by the Center for
Defect Physics, an Energy Frontier Research Center funded by the
U.S. DOE, Office of Science, Basic Energy Sciences.

\bibliographystyle{hapalike}
\bibliography{caari-paper-arxiv}

\end{document}